# XNAT-PIC: Extending XNAT to Preclinical Imaging Centers

## Authors

Sara Zullino[1], Alessandro Paglialonga[1], Walter Dastrù[1], Dario Livio Longo[2*], Silvio Aime[1]

## Affiliations

[1]Molecular Imaging Center, Department of Molecular Biotechnology and Health Sciences, University of Torino, Torino, Italy; Euro-BioImaging ERIC, Torino, Italy

[2]Institute of Biostructures and Bioimaging (IBB), Italian National Research Council (CNR), Torino, Italy

## ORCID iD's

Sara Zullino, https://orcid.org/0000-0003-3066-9357

## Corresponding author

*Dario Livio Longo, Institute of Biostructures and Bioimaging (IBB), Italian National Research Council (CNR), Via Nizza 52, 10126, Torino, Italy

Phone: +390116706473, Fax: +390116706487

email: dario.longo@unito.it

## Running title

XNAT-PIC: XNAT for Preclinical Imaging Centers

## Abstract

Molecular imaging generates large volumes of heterogeneous biomedical imagery with an impelling need of guidelines for handling image data. Although several successful solutions have been implemented for human epidemiologic studies, few and limited approaches have been proposed for animal population studies. Preclinical imaging research deals with a variety of machinery yielding tons of raw data but the current practices to store and distribute image data are inadequate. Therefore, standard tools for the analysis of large image datasets need to be established. In this paper, we present an extension of XNAT for Preclinical Imaging Centers (XNAT-PIC). XNAT is a worldwide used, open-source platform for securely hosting, sharing, and processing of clinical imaging studies. Despite its success, neither tools for importing large, multimodal preclinical image datasets nor pipelines for processing whole imaging studies are yet available in XNAT. In order to overcome these limitations, we have developed several tools to expand the XNAT core functionalities for supporting preclinical imaging facilities. Our aim is to streamline the management and exchange of image data within the preclinical imaging community, thereby enhancing the reproducibility of the results of image processing and promoting open science practices.

**Introduction**

Preclinical imaging deals with the visualization of small animals, such as mice and rats, for research purposes, in order to assay biological structures and activities *in vivo*, thus providing quantifiable, spatial and temporal information on healthy and pathological tissues down to both cellular and molecular level [1]. Importantly, because of its non-invasiveness, imaging is suitable for longitudinal studies of animal models in the fields of diagnostics, epidemiology and drug development [2].

Imaging research generates large amounts of data and information, mostly produced by computers and laboratory instruments. Consequently, it is of utmost important to provide appropriate data storage to ensure adequate management and organization to the research labs. Data loss not only implies losing processed images and the know-how gained, but also a waste of time and resources. Data sharing among research groups is another critical factor, particularly in scientific collaborations that involve many partners. In the last decade, several image repositories have emerged enabling the discovery of datasets from peer-reviewed publications or research studies in the life science domain, from biological imaging such as electron/fluorescence microscopy and high content screening, to biomedical imaging, such as Magnetic Resonance Imaging (MRI), Positron Electron Tomography (PET), Computed Tomography (CT) and Ultrasound (US) [3]–[5]. These resources can now be retrieved online by users through a browser user interface or programmatically by using Application Programming Interface (API).

For human population studies on biomedical imaging, large data repositories are routinely used by researchers and physicians all over the world. Among them, the Human Connectome Project [6] is a compilation of neural data, The Cancer Imaging Archive (TCIA) is a broad collection of cancer image data [7], the Alzheimer's Disease Neuroimaging Initiative (ADNI) is a shared catalogue of image data related to the Alzheimer's Disease [8] and the Open Access Series of Imaging Studies (OASIS) is a repository of magnetic resonance images [9].

The processing of large volumes of biomedical image datasets requires dedicated platforms and proper infrastructures. The Longitudinal Online Research and Imaging System (LORIS) is a flexible online system for data management devoted to multicenter studies that covers all the aspects from data acquisition from multiple sources to storage, processing, and dissemination [10]. Other examples of extensible data management platforms are The Human Imaging Database (HID) and the Collaborative Informatics and Neuroimaging Suite (COINS) for clinical neuroimaging studies [11], [12]. To tackle the needs of heterogeneous studies, the Medical Imaging Research Management and Associated Information Database (MIRMAID), a web accessible content management system for medical images was proposed [13]. More recently, Anastasopoulos et al. introduced Nora Imaging, an in-house web platform, to process medical imaging data derived from brain imaging studies [14].

The Extensible Neuroimaging Archive Toolkit (XNAT) is an imaging informatics system designed by the Neuroinformatics Research Group at the Washington University to manage images from several sources, to save data in a safe database, and to share data among authorized users [15], [16]. It was originally conceived to deal with data management of neuroimaging laboratories, but its increasing success has promoted its use in many other medical imaging fields. Over the years, XNAT has become a large project sustained by active user and developer communities. Indeed, many academic institutions and hospitals build their own data management and process infrastructure upon the XNAT system with the aim to extend its core features and provide new functionalities to meet the needs of researchers and physicians. An XNAT-based framework for managing large scale clinical studies was developed by Doran et al. along with in-house applications for data selection and uploading [17]. GIFT-Cloud is a medical image sharing platform built on top of XNAT 1.6 dedicated to GIFT-Surg, an international scientific project that develops novel imaging techniques for prenatal surgery [18]. Moreover, XNAT has been customized for automated quality assessment of retinal Optical Coherence Tomography (OCT) [19] and for collaborative research in human sleep medicine [20]. XNAT has also been used to promote multicenter reproducibility studies for radiomics [21]. The European Population Imaging Infrastructure (EPI2) provides an XNAT-based environment for the implementation of large, prospective epidemiological imaging studies, allowing for permanent and/or temporary storage of medical images. Furthermore, EPI2 develops state-of-the-art image analysis pipelines for high volume image processing [22], [23]. Lastly, Health-RI offers an XNAT-based service for research projects related to archive, view and process clinical imaging data [24], [25].

Animal population imaging allows researchers to test novel therapies and drugs on small animal models that will be eventually translated into the clinic. On the other hand, it suffers from the lack of standard tools to store, process, and share imaging data produced by scientists. Kain et al. have recently proposed the Small Animal Shanoir (SAS), an expansion of the Shanoir platform that was developed for data management dedicated to human neuroimaging repositories [26], [27]. SAS offers a cloud-based solution for exchanging data and analysis tools for small animal imaging studies in the framework of France Life Imaging (FLI).

Small animal imaging facilities are highly specialized centers that provide the research community, both academic institutions and industrial enterprises, with access to cutting-edge imaging technologies. Therefore, animal imaging centers have to deal with the complexity and the variety of preclinical trial datasets. The time needed to perform these studies and the relative costs are prompting imaging scientists to share image data in public repositories. Sharing data from independent research investigations, beyond saving resources, ensures the reproducibility of the experiments, aids the scientific community to get noticed and boosts collaborations among research institutions. The main difficulties arise from the complexity of the analysis and the scarcity of standard applications for sharing and processing preclinical images.

Vendors of preclinical imaging machinery use proprietary formats to save the acquired images. This make them are hard to handle by users with limited experience in data management and image processing. These drawbacks arise significant questions for multimodal image investigations especially in recording, curation, and processing of imaging data. Commercial softwares distributed by imaging device manufacturers provide the possibility to export proprietary raw image datasets to internationally recognized data standards, being Digital Imaging and COmmunications in Medicine (DICOM) [28] and Neuroimaging Informatics Technology Initiative (NIfTI) [29] the most popular ones. Notably, the storage of the parameters related to the image acquisition is poorly supported by NIfTI, as opposed to DICOM [30]. Despite DICOM being the internationally acknowledged standard for the interchange and management of medical images, some sections still require coherent semantics to guarantee consistent sharing of image data across several third-party products. In addition, the DICOM standard needs to be updated and expanded to support novel applications and incorporate emerging imaging technologies. In particular, at preclinical level new techniques have been developed such as Chemical Exchange

Saturation Transfer (CEST) MRI, Photoacoustic Imaging (PAI) and Optical Imaging (OI), so that standardization as well as *ad-hoc* DICOM attributes are desirable.

This work aims at the development of an XNAT-based platform that meets the needs of the preclinical imaging community. This system is named XNAT-PIC that stands for XNAT for Preclinical Imaging Centers. Our goal is double: we intend to provide the preclinical imaging researchers with tools to easily extract, import, and archive biomedical image data and with reusable processes for high-throughput extraction of quantitative features from raw image data, all within an XNAT environment. All the developments are free and open-source. We believe that XNAT-PIC will allow a streamlined exchange and reuse of image data among preclinical imaging facilities.

**Materials and Methods**

The Molecular Imaging Center (CIM) hosted at the Department of Molecular Biotechnology and Health Sciences of the University of Torino has deployed an XNAT instance devoted to preclinical imaging available at http://cim-xnat.unito.it. XNAT is a free and open-source Java based web application, exploiting the PostgreSQL database system. Users can personalize an instance and broaden its basic features to support their data and project management needs. The deployment presented in this paper is running on XNAT 1.7.6, using Apache Tomcat 7, Oracle JDK 8, PostgreSQL 11, and Ubuntu 20.04 LTS Operative System.

XNAT natively supports multiple imaging modalities, such as MRI, PET, CT, and US with the possibility to extend XNAT datatypes to other imaging methods. XNAT comes with a built-in DICOM image management application that also allows to store images in NIfTI format. It offers key functionalities, for instance importing and downloading images in multiple formats, archiving, and distributing data, and setting data protection and accessibility. In XNAT, users can either store the raw or processed images on local disks or transfer them through the network to a DICOM C-STORE Service Class Provider (SCP), for instance a Picture Archive and Communications Systems (PACS) or another XNAT deployment. Moreover, XNAT comes with a Java-based viewer to access and inspect the images in the archive. This viewer can be customized with plugins to integrate supplementary functionalities specific to the image type of interest.

XNAT for Preclinical Imaging Centers (XNAT-PIC) has been developed to meet the needs of the preclinical imaging community. In particular, XNAT-PIC consists of a suite of tools aimed at converting raw MR image series to DICOM standard, uploading to XNAT and processing large image datasets. This workflow is based on the steps schematically depicted in Figure 1:

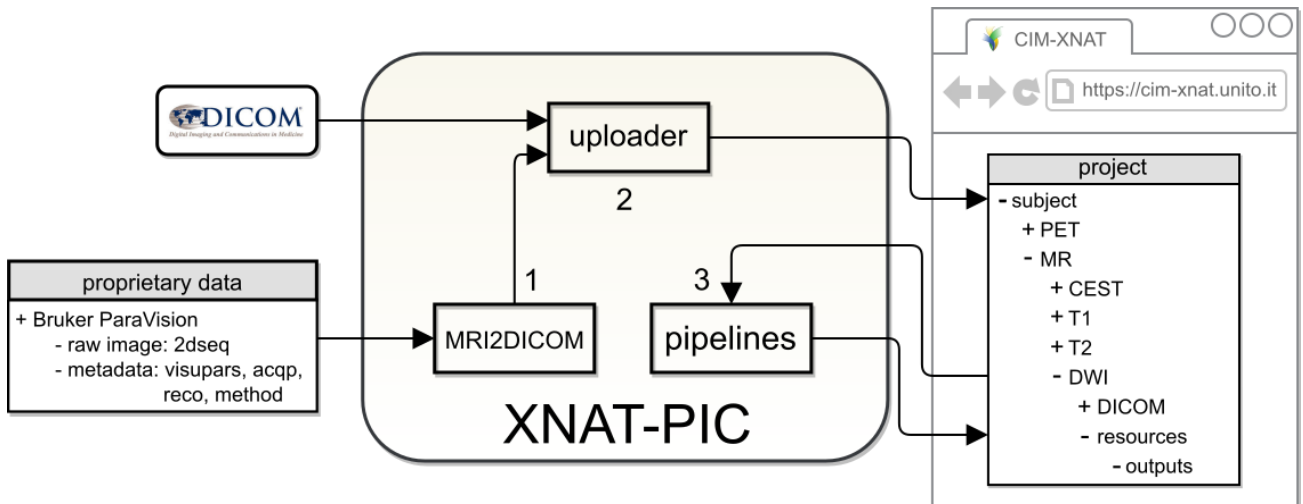

**Figure 1:** Schematic workflow of image archiving and processing. XNAT-PIC is a suite of tools aimed at facilitating the management and the analysis of preclinical image datasets.

1. MRI2DICOM, a MR image converter from ParaVision® (Bruker, Inc. Billerica, MA) file format to DICOM standard.
2. XNAT-PIC Uploader to upload large, multimodal image datasets in DICOM standard to XNAT.
3. XNAT-PIC Pipelines for processing single or multiple subjects in an XNAT project.

MRI2DICOM is a free and open-source tool built in Python 3.7.6 downloadable at https://github.com/szullino/XNAT-PIC [31]. The application uses the numpy 1.15.4 and Pydicom 1.2.1 libraries to handle DICOM files [32]–[35]. MRI2DICOM has been designed for and tested on different several of ParaVision®, such as 5.1, 6.0.1 and 360.1.1. It accesses the ParaVision® data structure, deciphers the binary file (*2dseq*) containing the image and parses the acquisition parameters stored in different files (*visupars*, *method*, *reco*, and *acqp*) into Python dictionaries [36]. Lastly, it saves all the relevant information into the DICOM header and image set field, according to the MRI acquisition protocol.

XNAT-PIC Uploader is built in Python 3.7.6. The communication with XNAT is possible through xnatpy 0.3.22, a new and open-source XNAT Python client, and pyAesCrypt 0.4.3 Python library to encrypt the files containing the XNAT login credentials [37], [38]. PyInstaller 3.5 has been used to bundle the Python applications and all its dependencies into a single package to run MRI2DICOM and XNAT-PIC Uploader as a stand-alone executable, in both Windows and Linux environments [39].

XNAT-PIC Pipelines have been designed to process DICOM images stored in XNAT, extract quantitative information, and produce parametric maps. These pipelines have been built on top of pre-existing image analysis scripts developed in MATLAB R2020b (The MathWorks, Inc., US) by our group. XNAT-PIC Pipelines can be of two types: *subject level* or *project level pipeline*. *Subject level pipelines* consist of a bash script that invokes MATLAB to perform the computation. A different approach has been used for *project level pipelines*, i.e. to process multiple subjects within the same project. A Python 2.7 virtual environment has been created and the following packages have been installed: i) the MATLAB Engine API for Python to run MATLAB scripts within a Python session [40], ii) pyxnat-1.2.1.0.post3 that facilitates scripting interactions with the XNAT database [41], [42], iii) Requests 2.23.0 that makes Hypertext Transfer Protocol (HTTP) requests to communicate with XNAT extremely simple [43]. Finally, the *Mask Average Pipeline* has been developed to calculate the mean value in a Region of Interest (ROI) of a parametric map generated from processing. This pipeline runs on Python 3.8.3 and uses the following libraries: numpy 1.18.5 [44] dcmrtstruct2nii 1.0.19 to export DICOM Radiotherapy Structure Set (RTSTRUCT) to NIfTI mask [45], the image processing library opencv-python 4.4.0.40 [46], and nibabel 3.1.1 for manipulating NIfTI images [47].

## Results

**MRI2DICOM Converter**
Upon launching, XNAT-PIC offers multiple functions as shown in Figure 2A: users can use MRI2DICOM to convert the binary data to DICOM standard or, given the data already in DICOM, users can use XNAT-PIC Uploader to import the MR image sessions to XNAT. If MRI2DICOM has been selected by the user, the converter needs to know the directory of the project in ParaVision® format to start the conversion (Figure 2B). Once the process is completed, a new folder containing the project in DICOM standard will be created at the same location as the original one in raw format.

Commercial softwares in preclinical imaging poorly conform to standards regarding data storage and each vendor uses a proprietary format for its data. In order to deal with emerging

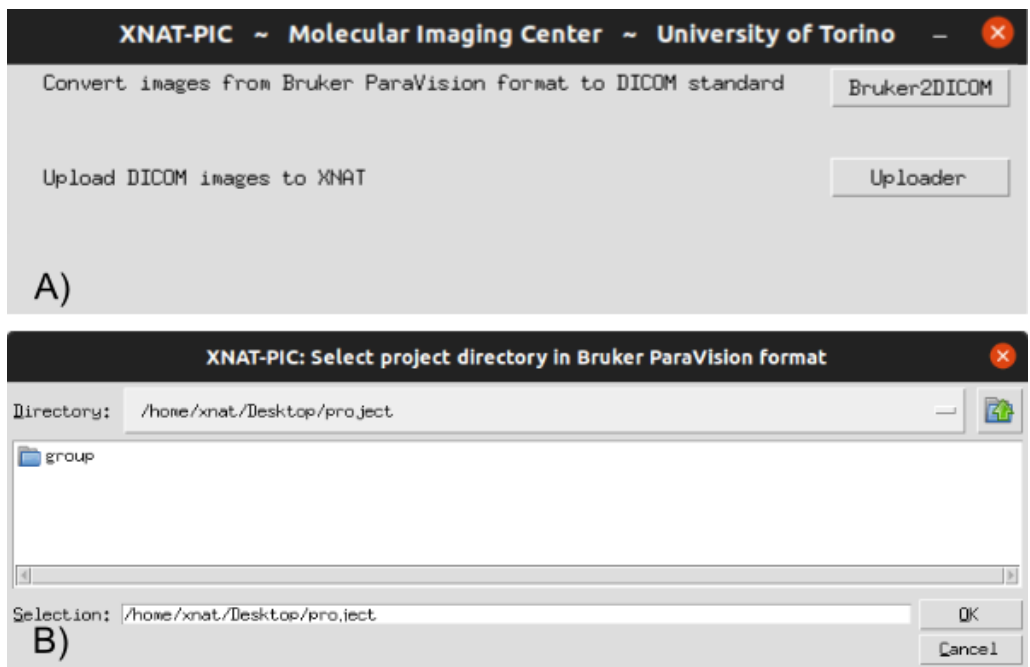

**Figure 2:** Snapshots of the XNAT-PIC application. A) Upon launch, users can choose between converting raw image data to DICOM standard or upload pre-existing DICOM images to XNAT. B) If the converter has been selected, MRI2DICOM allows the user to browse to the directory of the project in ParaVision® format.

MRI techniques, a novel set of vocabularies can be introduced as private DICOM attributes specifically devoted to these modalities [48]. For example, new dictionary items have been dynamically added to the 'standard' DICOM dictionary to describe CEST-MRI datasets. The full list of DICOM attributes is available in section 6 of DICOM standard [49]. To avoid conflicts of private attributes from different implementers, a block in the (0061, 0010) space has been reserved for this specific use. Table *1* shows the private DICOM dictionary that has been developed for CEST-MRI acquisitions along with the corresponding entry gathered from the *method* file of ParaVision® 360.1.1. The Value Multiplicity (VM) of an attribute defines the number of values contained in that attribute, while the Value Representation (VR) describes the data type and format of each DICOM attribute.

| Attribute Name | Tag | VM | VR | Definition | PV360 |
|---|---|---|---|---|---|
| **CEST Parameters Creator** | (1061,0010) | 1 | LO | Creator of the parameter set | OWNER |
| **Chemical Exchange Saturation Method** | (1061,1001) | 1 | LO | Method of the | Method |
| | | | | acquisition sequence | |
| **Saturation Type** | (1061,1002) | 1 | LO | Types of saturation transfer mechanisms | PVM_SatTransType |
| **Pulse Shape** | (1061,1003) | 1 | LO | Shape of the saturation pulse | PVM_SatTransPulseEnum |
| **B1 Saturation** | (1061,1004) | 1 | DS | B1 field of the RF pulse peak amplitude in µT | PVM_SatTransPulseAmpl_uT |
| **Pulse Length** | (1061,1005) | 1 | DS | Length (duration) of the saturation pulse in ms | PVM_SatTransPulse |
| **Pulse Number** | (1061,1006) | 1 | DS | Number of saturation pulses | PVM_SatTransNPulses |
| **Interpulse Delay** | (1061,1007) | 1 | DS | Interval in ms between pulses in a pulsed saturation scheme | PVM_SatTransInterpulseDelay |
| **Saturation Length (ms)** | (1061,1008) | 1 | DS | Pulse Length × Pulse Number | PVM_SatTransModuleTime |
| **Readout Time (ms)** | (1061,1009) | 1 | DS | Time needed for readout | computed |
| **Pulse Length 2 (ms)** | (1061,1010) | 1 | DS | Length of the second saturation pulse in an uneven irradiation scheme | PVM_SatTransPulseLength2 |
| **Duty Cycle** | (1061,1011) | 1 | DS | Fraction of one period where the pulse is active | computed |
| **Recovery Time (ms)** | (1061,1012) | 1 | DS | Time between the end of the readout and the beginning of | computed |

| | | | | the next saturation (Repetition Time - Pulse Length – Readout Time) | |
|---|---|---|---|---|---|
| **Measurement Number** | (1061,1013) | 1 | DS | Number of frequency offsets | PVM_NSatFreq |
| **Saturation Offset (Hz)** | (1061,1014) | 1 | DS | Frequency offsets list in Hz | PVM_SatTransFreqValues |
| **Saturation Offset (ppm)** | (1061,1015) | 1 | DS | Frequency offsets list in ppm | Computed if PVM_SatTransFreqUnit = unit_hz, and viceversa |

**Table 1:** Private DICOM attributes for CEST-MRI modality gathered from the ParaVision® *method* file. To avoid conflicts of private tags from different vendors, a block in the (0061, 0010) space has been reserved for this specific imaging modality. VM = Value Multiplicity. VR = Value Representation. LO = Long String. DS = Decimal String.

Preclinical investigations may cover a wide variability of studies addressing specific scientific questions. Therefore, the data organization is usually tailored to the study of interest and needs to be matched to the XNAT data hierarchy. The capability of XNAT to manage different imaging data structures has then been extended, including longitudinal studies (several timepoints), treatments procedures (control and treated groups) and drug administrations (doses). The flexibility in managing several experimental conditions has been achieved by developing specific custom variables. Besides the default variables, XNAT users can specify an unlimited number of personalized variables. These custom variables can be reused within other projects, allowing for inter-project standardization. For longitudinal studies, three sets of custom variables have been created: the term "group" refers to the treatment protocol and can have two possible values, "treated" or "untreated" (i.e., control group), "timepoint" is related to the timing of the acquisitions (i.e. before ($t_0$) and after ($t_1$, $t_2$) a treatment, or simply different timepoints) and "dose" refers to a specific drug dosage in therapeutic treatment investigations (Figure 3). Notably, XNAT-PIC users can adjust the custom variables accordingly to their data structure, provided that the custom variables are at most three and their naming conventions in XNAT are respected.

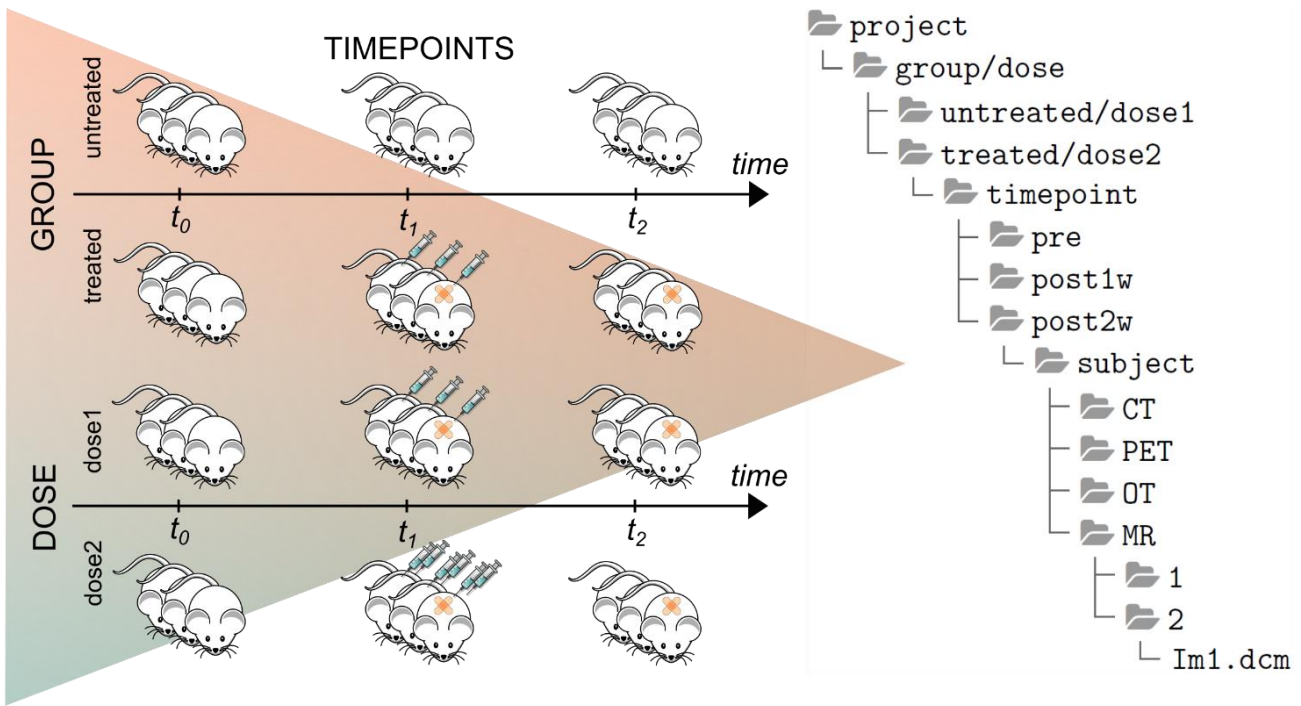

**Figure 3:** Schematic representation of the custom variables implemented in XNAT to match a typical data organization in preclinical longitudinal studies: "group" refers to treated and untreated mice; "timepoints" is related to the timing of the acquisition: t0 = pre, before treatment; t1 = post1w, 1 week after; t2 = post2w, 2 weeks after; "dose" refers to amount of drug (dose1 and dose2, respectively) used in this specific experiment.

**XNAT-PIC Uploader**

Upon conversion to DICOM format, the image dataset can be uploaded to XNAT. XNAT-PIC Uploader is designed to perform single subject uploading or batch uploading. It needs the XNAT webpage address and the login details (Figure 4A); then users can generate a new project or choose an existing one in the list (Figure 4B), navigate to the directory of interest and entry the number of custom variables (Figure 4C). A pop-up message notifies the user once the dataset is successfully imported to XNAT. The original raw images can be eventually uploaded to the database as project level resources.

**Figure 4:** Snapshots of XNAT-PIC Uploader. A) Accessing XNAT requires the login details, such as the XNAT web address, username, and password; B) Users can create a new project or click the drop-down menu to select an existing project in the list; C) Custom variables can be entered by typing the number of variables in the 1 to 3 range that corresponds to number of folders in the project directory.

Preclinical studies may imply the use of several imaging modalities that further increase the complexity of data management, since the same patient can typically undergo several examinations with different imaging instrumentation. For instance, studies that investigate tumor metabolism and acidosis may require the combined use of 18-fluorodeoxyglucose (FDG) PET technique for measuring tumor glucose uptake and CEST-MRI pH imaging for assessing tumor acidosis [48], [50]. The XNAT-PIC Uploader has been designed to upload image datasets of different modalities to XNAT such as MRI, PET, CT, and US, allowing users to efficiently manage large, multimodal imaging studies.

The XNAT-PIC Uploader has been tested and validated on several studies, for instance including data gathered from CEST-MRI experiments, in particular GlucoCEST imaging for assessing tumor metabolism following glucose injection [51], [52]. Figure 5 shows a snapshot of the project and subject webpage in XNAT where the custom variables "group" and "timepoints" have been used to describe the GLINT project.

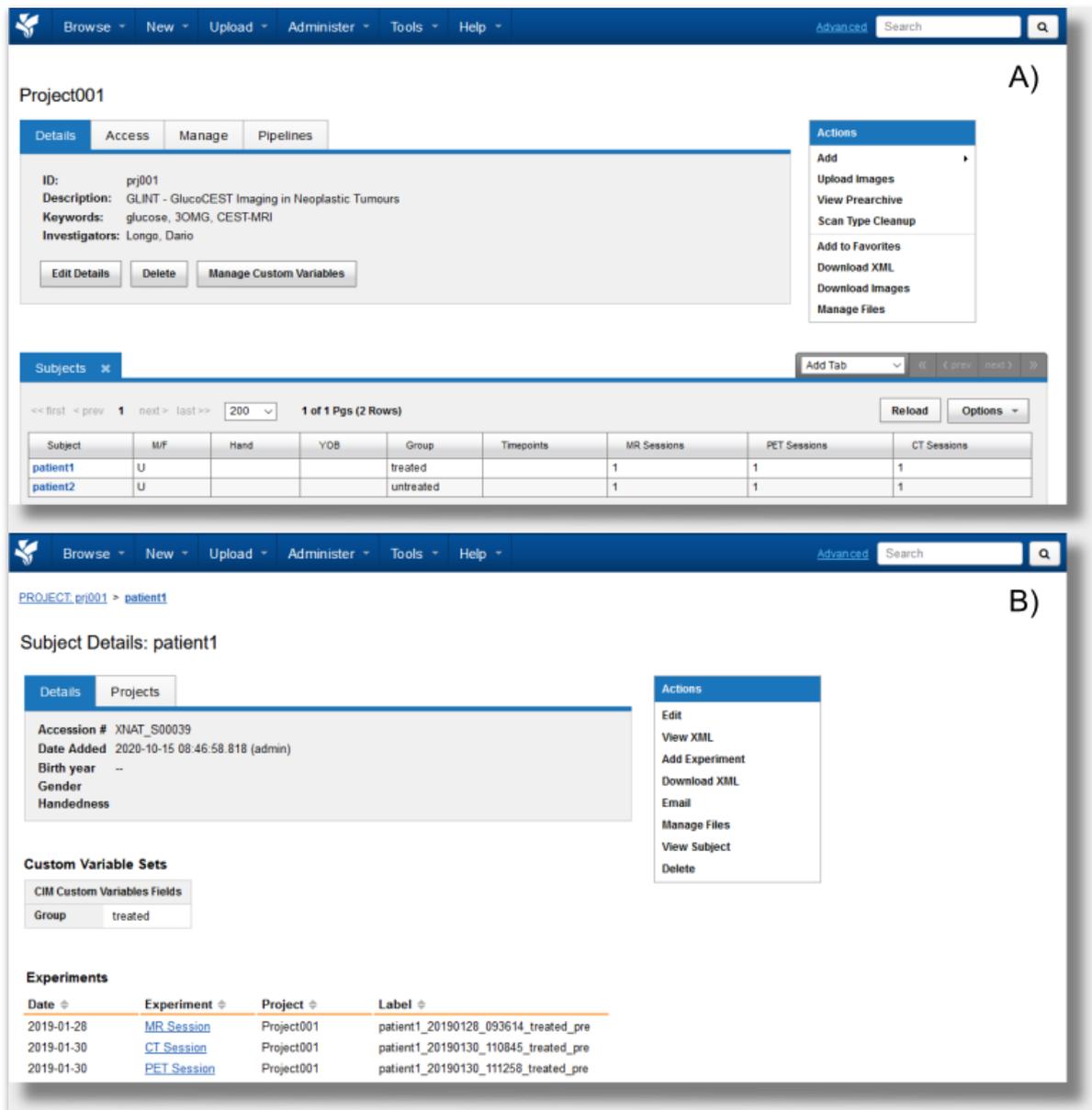

**Figure 5:** Snapshots of the project (panel A) and subject (panel B) webpage in XNAT with multiple image sessions (MR, CT, and PET, respectively). The custom variable "group" describes the patient status (treated or untreated) and is shown at each level of the XNAT data hierarchy, while the custom variable "timepoint" referring to the timing of the treatment administration is displayed at the subject level (panel B) in the Label field.

**XNAT-PIC Pipelines**

XNAT's core activity is the execution of pipelines. A pipeline is defined as a sequence of steps. Each step of the pipeline invokes an external application, a shell script, or an executable for processing the data. The pipeline is executed by the XNAT Pipeline Engine, a Java-based framework efficiently integrated within the XNAT platform [15]. The pipeline engine works with simple data flows on a step-by-step basis and can perform computational tasks on project data. The workflow is defined in an eXtensible Markup Language (XML) document named *pipeline descriptor* and the executables are defined in another XML document named *resource descriptors*.

Pipelines need to be enabled through the *Pipelines* tab in the project webpage. By clicking the *Add More Pipelines* button, the *Pipelines* tab will display a table of pipelines currently installed in XNAT. Users can add and configure a pipeline to be registered for the project.

XNAT is currently designed to run pipelines at subject level only. This is not feasible if one needs to process large scale image datasets. The urgency was therefore to scale up this approach and create pipelines to process an entire project. Our aim was to ensure the same user experience as originally provided by XNAT when launching a pipeline for all the subjects in a project. The main idea was to build a project level pipeline that can be initiated from any subject within the selected project and iterated over all the subjects in that project. MATLAB R2020b scripts previously developed by our group have been used to build pipelines in XNAT and process a variety of MRI acquisitions, such as $T_1/T_2$ mapping and Diffusion Weighted Imaging (DWI). Two approaches have been designed and tested to process these MR images at subject and project level, respectively.

*Subject Level Pipeline*

An XNAT pipeline can be launched via the *Run Pipeline* tool in the *Actions Menu* in the MR Session webpage. A pop-up window reports a list of all the pipelines available for that project. This list contains both the pipeline processing only the current subject and the pipeline processing all the subjects in that project. Figure 6 shows a screenshot of the pop-up windows where users can select between two pipelines: *process_T2* processes the $T_2$ map of the current subject, while *process_T2_project* processes the $T_2$ maps of all the subjects in the same project. Upon selection of the subject level pipeline, users are then asked via web browser to tick the scan number to be analyzed. The *pipeline descriptor* contains XML instructions to take this user input, create the working directory, download the DICOM files corresponding to the scan of choice, upload the results of the analysis back to XNAT in a dedicated *resource* folder corresponding to the scan of interest and call the *resource descriptor*. The *resource descriptor* invokes a bash script that passes the DICOM files directory to MATLAB running in background to process those images.

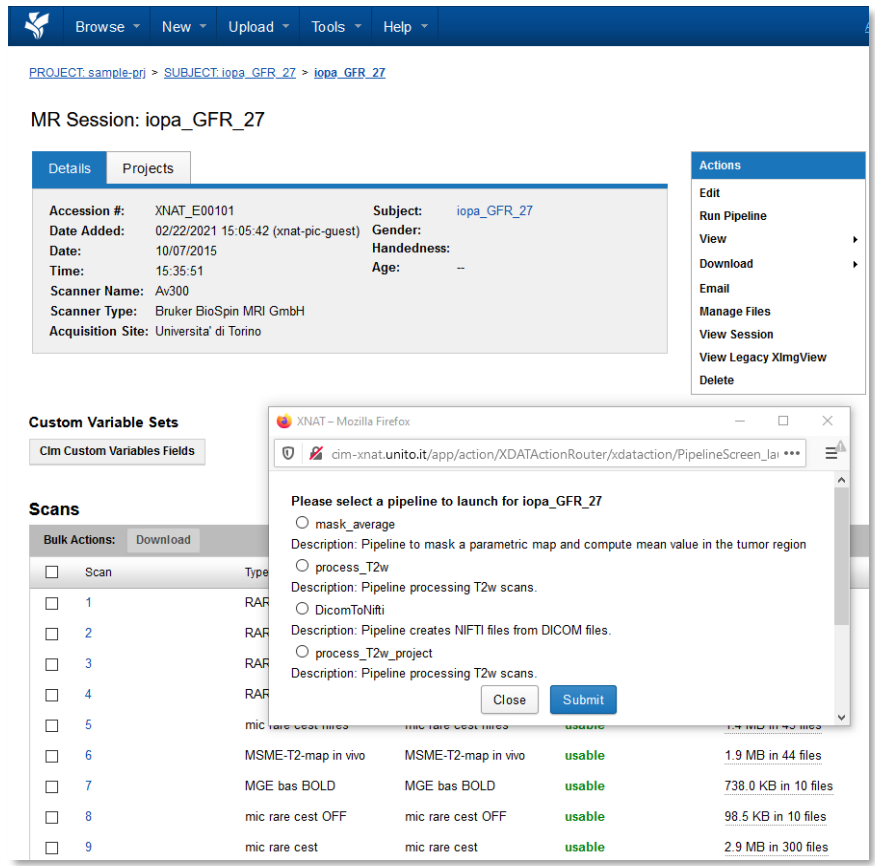

**Figure 6:** Screenshot of the pop-up windows with the list of pipelines available for this project. Users can select between two pipelines: *process_T2* processes the $T_2$ map of the current subject, while *process_T2_project* processes the $T_2$ maps of all the subjects in the same project.

### *Project Level Pipeline*

Users may need to process all the subjects of a project in a single run. Therefore, they can select a project level pipeline from the *Run Pipeline* pop-up menu.

In this case, the pipeline descriptor contains XML instructions to create the working directories needed for the processing and call the *resource descriptor*. The *resource descriptor* invokes a bash script running a Python wrapper consisting of a sequence of Python scripts in order to:

1. Search for the images within the project that have a specific DICOM attribute describing the acquisition protocol of the scans to be processed.
2. For each subject, create a local folder and then download the corresponding DICOM images.
3. Loop over the subjects and locally execute the MATLAB scripts.
4. Upload the results back to XNAT in a *resource* folder created by the pipeline at subject level.

In order to run XNAT-PIC pipelines, two scenarios are possible: i) generic users can register to our CIM-XNAT instance, upload their image datasets and add the pipelines to their own projects (See: Adding Pipelines To Your Project: https://wiki.xnat.org/documentation/how-to-use-xnat/adding-pipelines-to-your-project); ii) XNAT Admins can download XNAT-PIC pipelines from https://github.com/szullino/XNAT-PIC-Pipelines, install and register the pipelines in their own XNAT instance (See: Installing Pipelines in XNAT:

https://wiki.xnat.org/documentation/xnat-administration/configuring-the-pipeline-engine/installing-pipelines-in-xnat).

Table 2 lists the processing pipelines currently installed on our XNAT deployment and available for download.

| Name | Description |
|---|---|
| Process DWI | Pipeline processes DWI map |
| Process DWI project | Pipeline processes all DWI maps in a project |
| Process T1w SR | Pipeline processes $T_1$ Saturation Recovery map |
| Process T1w SR project | Pipeline processes all $T_1$ Saturation Recovery maps in a project |
| Process T2 | Pipeline processes $T_2$ map |
| Process T2 project | Pipeline processes $T_2$ maps in a project |
| Mask Average | Pipeline computes a mean value in the ROI |

**Table 2:** Processing pipelines currently installed on our CIM-XNAT instance (http://cim-xnat.unito.it) and available for download at https://github.com/szullino/XNAT-PIC-Pipelines.

In Figure 7, the workflow for processing DWI acquisitions is schematically presented. Post-processed image data and other files are uploaded back to XNAT in the resource folder created at subject level or scan level, according to the full project or standard subject pipeline respectively, and accessible by the user through a *Manage Files* console. The resource folder can contain several subfolders, each of them populated with several files, such as parametric images in several file formats (i.e., NIfTI), MATLAB workspace, log files, and many others.

**Mask Average Pipeline**

Following the processing, preclinical imaging users are usually interested in performing some simple statistics on the processed images. We have therefore developed a pipeline that applies a mask to a parametric map and compute the mean value in the ROI. To do so, the XNAT OHIF Viewer 2.0 Plugin has been installed on our XNAT instance [53]. The plugin is an integration of the Open Health Imaging Foundation (OHIF) viewer into XNAT [54]. Users can create contour-based, such as DICOM RTSTRUCT and Annotation Image Markup (AIM) as well as mask-based DICOM Segmentation (DICOM SEG) ROI Collections, and import/export them to ROI Collection Assessors in XNAT (Figure 9). Once the ROI has been drawn on the $T_2$-weighted image and saved to XNAT, the *Mask_Average* pipeline can be launched from the *Run Pipeline* tool. The *pipeline descriptor* contains XML instructions to take the scans corresponding both to the T$_2$-weighted acquisition and the parametric map in the resource folder, download the morphological image in DICOM standard and the parametric image in NIfTI format corresponding to the user selection, download the ROI Collection Assessors, create the working directory, upload the results of the analysis back to XNAT in a dedicated *resource* folder corresponding to the scan of interest and call the *resource descriptor*. The *resource descriptor* invokes a bash script passing both the T$_2$-weighted DICOM image and the DICOM RT-STRUCT directories to a Python code to be converted into a NIfTI mask by the Python package dcmrtstruct2nii [55]. The same bash script passes the resulted mask to a Python script that computes statistical calculations in the ROI and uploads the results back to the database as XNAT resources (Figure 8). A typical output of the *Mask_Average* pipeline is shown in Figure 10. The XNAT-OHIF plugin is used to draw the ROIs corresponding to the mouse kidneys on the $T_2$-weighted I image (Figure 10A). The full $T_2$ map obtained from the *process_T2* pipeline along with the relative $T_2$ map after masking are shown in Figure 10B and C, respectively. A text file with the result of the calculation is depicted in Figure 10D.

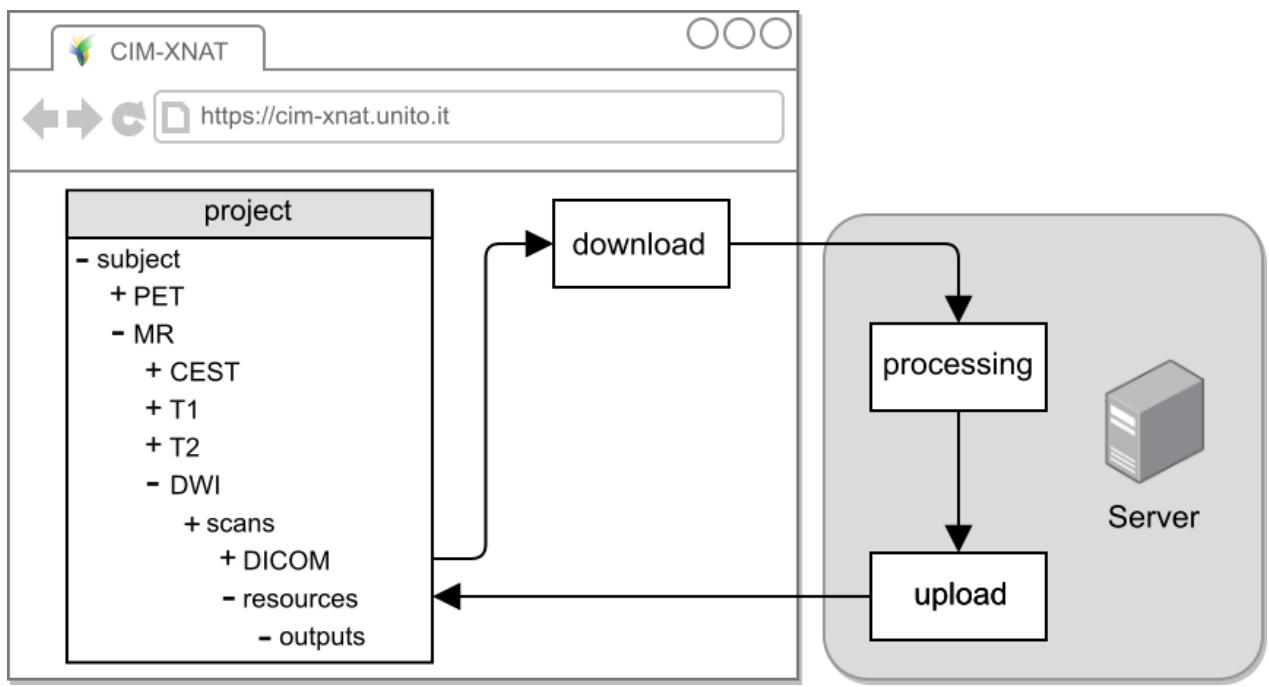

**Figure 7:** Schematic workflow of the *process_DWI* pipeline which retrieves, downloads, and processes the MRI diffusion scan. The output files (text file, log file, NIfTI images and MATLAB workspace) are then uploaded back to XNAT under the corresponding subject, experiment, and scan. This workflow is iterated when the project level *process_DWI_project* pipeline is launched to process all the MRI diffusion scans contained in the project; then, the resource folder is created at subject level.

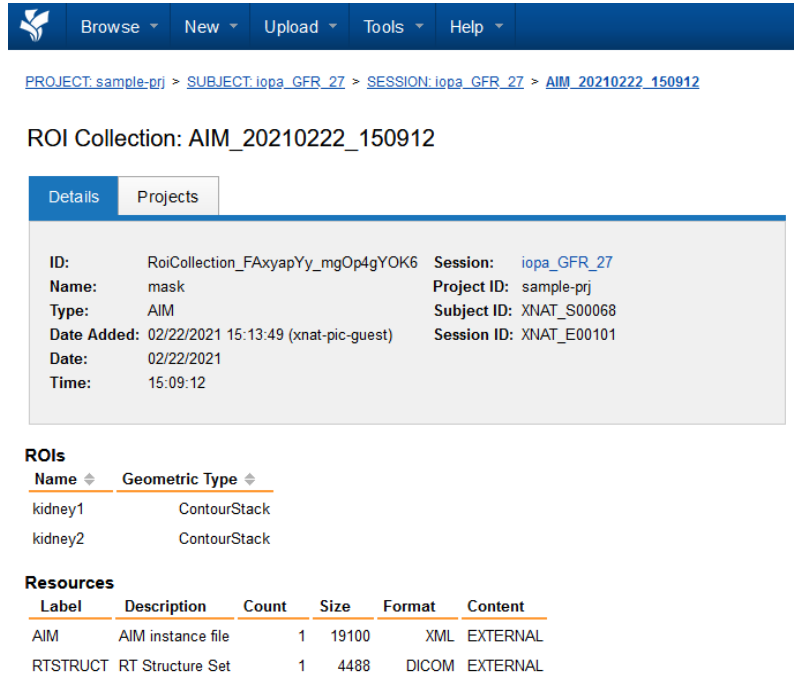

**Figure 8:** Snapshot of the ROI Collection Assessor webpage in XNAT. The mask represented here is composed by two ROIs, tumor1 and tumor2 respectively, and is saved in the DICOM RT-STRUCT file that needs to be converted in NIfTI format before usage.

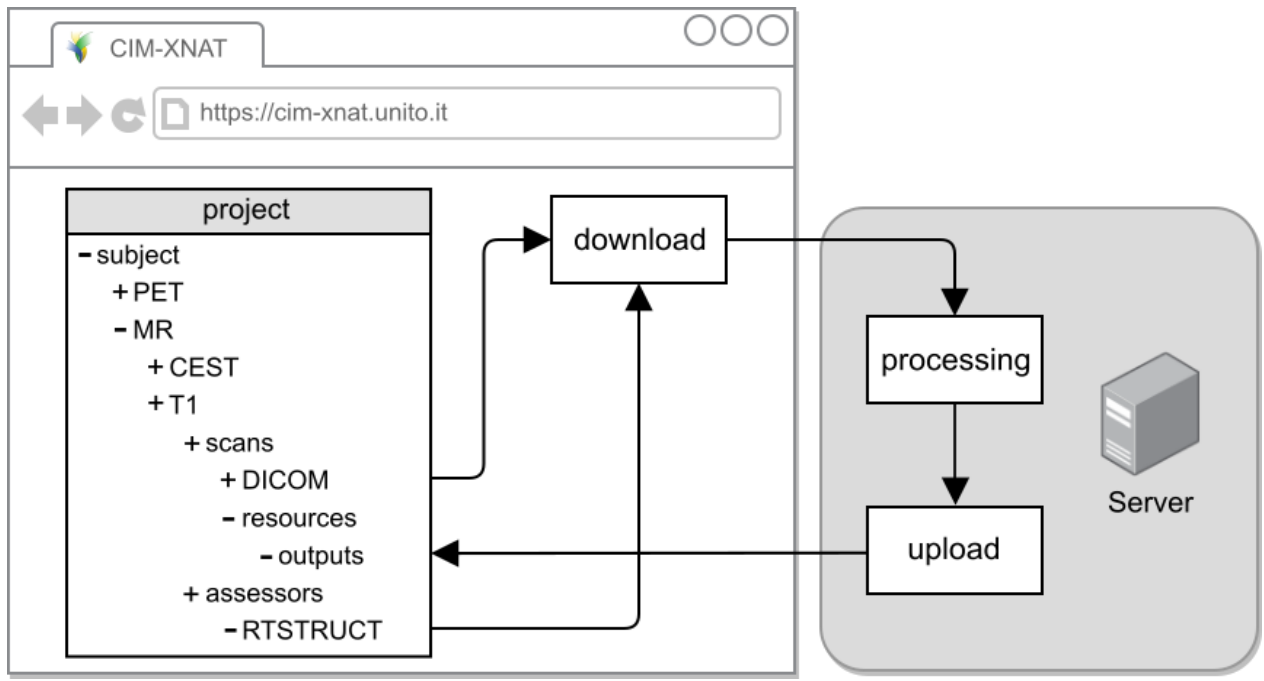

**Figure 9:** Schematic workflow of the *Mask_Average* pipeline that applies a mask to a parametric map in NIfTI format, computes the mean value in the ROI and uploads the masked parametric map and other output files back to XNAT.

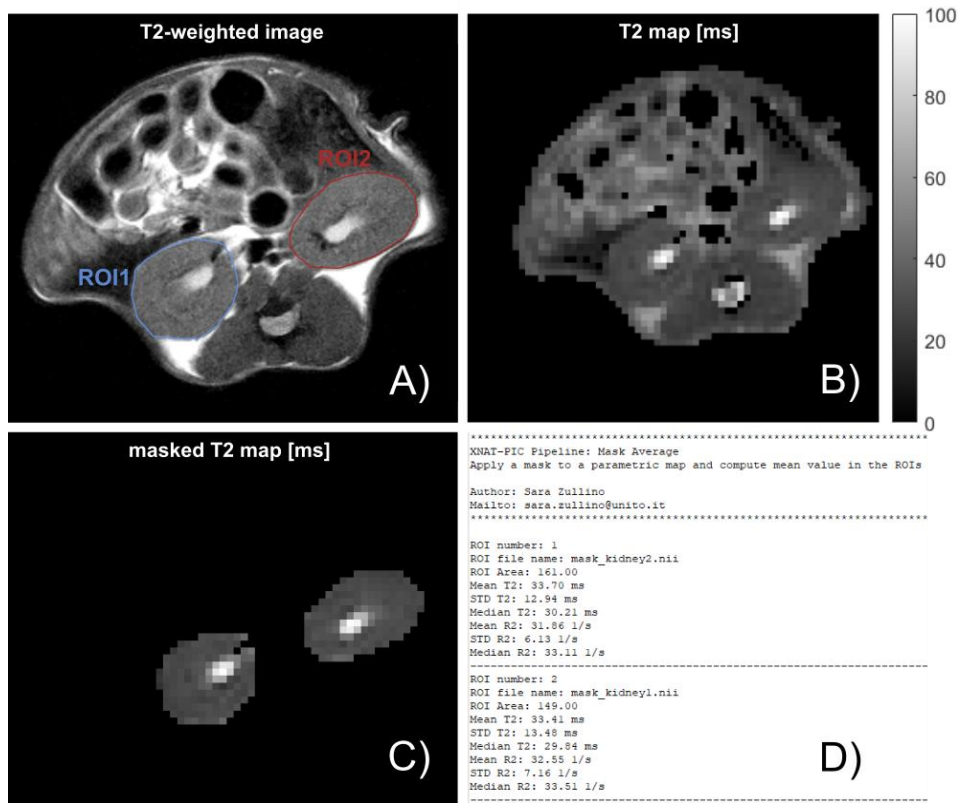

**Figure 10:** Outputs generated from the *Mask_Average* pipeline. A) Reference anatomic $T_2$-weighted axial image showing kidneys and corresponding ROIs. B) $T_2$ map obtained from *process_T2* pipeline. C) The mask is applied to the obtained $T_2$ map showing only the kidneys. D) Output text file reporting statistics for each ROI.

**Discussion**

Preclinical imaging facilities are usually equipped with a variety of imaging devices, each of them yielding large amounts of data usually stored in several workstations. Storing and sharing preclinical data in a safe, fast, and reliable manner is therefore imperative. Furthermore, image analysis is of utmost importance in preclinical applications to unravel the physiological mechanisms of the disease process or investigate the response to a new therapeutic treatment. Despite the urgent needs, few efforts have been done so far to manage preclinical imaging studies and ensure reproducibility through highly standardized image processing methods.

In this work we have developed XNAT-PIC, a free and open-source application consisting of a MR image converter from proprietary file format to DICOM standard, an uploader to import large and multi-modal imaging studies to XNAT and a catalogue of pipelines for processing preclinical image datasets. MRI2DICOM is a MR image converter from ParaVision® file format to DICOM that takes into account emerging MRI methods such as CEST imaging. A controlled vocabulary has been created for the first time to describe CEST-MRI acquisitions. The entries are extracted from the raw image files and used to feed a new, private DICOM dictionary specifically created for this modality. These DICOM attributes can be then easily accessed and reused by external applications for post-processing. XNAT-PIC Uploader has been designed to overcome an XNAT limitation regarding data upload. There are several options for uploading image sessions in XNAT: the XNAT Desktop Client and the XNAT Upload Assistant are stand-alone applications that can be installed for uploading and downloading data, while the XNAT Compressed Uploader is a tool that runs in the XNAT browser and does not require installation. Other options are the DICOM C-STORE Service Class Provider that can send data directly to the XNAT server and the Representational State Transfer (REST) API import service, the latter not applicable by non-expert users. To our knowledge, none of these methods is capable of uploading multiple subjects of different imaging modalities. The XNAT-PIC Uploader offers the possibility to upload several subjects screened with different modalities (i.e., MRI, PET, CT, and US) related to the same project in a single run through a user-friendly graphical interface. The imaging modalities that can be uploaded using the XNAT-PIC Uploader are the ones currently supported by XNAT, such as MRI, CT, PET and US. Future work will be needed to include data types related to imaging technologies that are popular in preclinical research, such as PAI and OI. Indeed, while PAI is labeled as US modality in DICOM standard, OI is still referred to Other (OT). At present, the list of DICOM modalities is available here [56]. Lastly, XNAT-PIC includes a collection of processing methods routinely used in preclinical imaging such as $T_1$ and $T_2$ mapping and DWI, with the aim to extend the offer of XNAT pipelines to other MRI techniques, firstly the emerging CEST-MRI imaging. Likewise, our intent is to improve the reproducibility of preclinical research and guarantee standardized image processing procedures by offering a free service to integrate custom-built analysis pipeline in the XNAT-PIC workflow upon user request.

The Small Animal Shanoir is another preclinical research solution developed to manage and process imaging datasets. It offers a platform to store and distribute the data, manage the metadata associated to the study, and process the images on high performance systems. In addition, the possibility to integrate custom-made processing pipelines and algorithms into SAS provides the reproducibility and accessibility of the research outputs. However, these services are subject to fees for both storage and processing. Interestingly, SAS architecture is based on micro-services, allowing SAS to function as a combination of independent structures that can be easily updated with new features or scaled up. Current, available micro-services in SAS comprise Dicomifier, a generic and open-source Bruker to DICOM or NIfTI converter, and applications to extract $T_1$ and $T_2$ relaxation times from MR image data. Recently, other commercial data management systems have been also introduced in preclinical imaging. The Small Animal Big-data warehouse Environment for Research (SABER) supports preclinical workflow and promotes data sharing, although this platform is unavailable in internet. Flywheel is a commercial data management system to store research data of different imaging modalities in a centralized archive, thus improving productivity and collaboration in life science research [57], [58].

The urgent need to develop infrastructures and services to support sharing and reusing of scholarly data motivated the establishment of the Findable, Accessible, Interoperable and Reusable (FAIR) Data Principle [59]. These rules are necessary to govern the scientific data management and stewardship and are applicable to several entities, such as industry, academia, scientific publishers, and funding agencies. The principles may serve as guidance for stakeholders willing to strengthen their data reusability. Unlike similar initiatives committed to human scholar, the FAIR Principles are intended to make the data discoverable and readable by both individuals and machines, therefore supporting their reuse [59], [60]. XNAT-PIC was developed in the framework of the demonstrator projects under the auspices of EOSC-Life, a European Union's Horizon 2020 research and innovation programme. XNAT-PIC is supported by the Multi-Modal Molecular Imaging Italian Node (MMMI) of Euro-BioImaging ERIC, a research infrastructure that offers open access to the most advanced imaging technologies, training and data services in biological and biomedical imaging [61]. In this scenario, XNAT-PIC acts as a first step towards data FAIRification in terms of Accessibility and Reusability. Generally, the data stored in an XNAT-based system is accessible upon authentication and authorization procedure: only trusted users have rights to access, manipulate and work with images. Data, tools and processing pipelines developed in XNAT-PIC are fully reusable as they are distributed under GNU General Public License v3 or any later version and available on GitHub. Some work needs still to be done regarding Findability and Interoperability. To date, the possibility to associate the dataset stored in XNAT with a Digital Object Identifier (DOI) or Persistent

Identifier is missing, preventing the data to be found by both humans and computers. In addition, machine-readable metadata that are necessary for data discovery are still needed. The European Open Science Cloud (EOSC) is an EU-funded project based on FAIR principles whose goal is to provide a public data repository compliant to open science principles. EOSC aims at providing "all researchers in Europe with seamless access to an open-by-default, efficient and cross-disciplinary environment for storing, accessing, reusing, and processing research data supported by FAIR data principles" as stated in The Vienna Declaration on the European Open Science Cloud [62]. Therefore, all the research materials that relate to scholarly data must be turned into FAIR. This includes the raw material, such as imaging dataset, as well as the tools, workflows, and pipelines needed to process the data, allowing to extract and quantify the information. The FAIR revolution also involves the standards, metadata, and ontologies that are necessary to provide significance to both the data itself and any complementary material.

Our future plan is to deploy a federated XNAT portal to collect preclinical imaging data from local XNAT installations and make them available to a broader community. The imaging community will largely benefit from this free, cloudified service, since it will enable users to discover image datasets normally not accessible, promoting the free exchange and reuse of data and ensuring higher standards of reproducibility of the experiments.

## Conclusion

While the basic XNAT deployment serves as a system for safely accessing, archiving, and processing clinical imaging studies, XNAT-PIC widens its core features in several ways to support preclinical imaging facilities. Our aim is to overcome the current limitations that arise from the management and the storage of preclinical imagery, thereby facilitating the analysis of biomedical image data.

The advantage of this approach relies in the capability to interface with several imaging modalities, including emerging imaging techniques such as CEST-MRI, manage different preclinical investigation protocols and easily process preclinical image data. We believe that such a workflow may be of interest for preclinical imaging centers, thus allowing the scientific community to efficiently store, process and share biomedical imaging data.

## Acknowledgements

The authors gratefully acknowledge Alessandra Viale (Euro-BioImaging ERIC Med-Hub, Torino) for her encouragement and support in the realization of the project; Stefan Klein,

Hakim Achterberg and Marcel Koek (Biomedical Imaging Group Rotterdam, Erasmus Medical Center, Rotterdam) for many fruitful discussions.

## Declarations

**Funding** (information that explains whether and by whom the research was supported)

This project has received funding from the European Union's Horizon 2020 research and innovation programmes under grant agreement No 824087 EOSC-Life, No 654248 CORBEL, No 667510 GLINT. The Italian Ministry for Education and Research (MIUR) is gratefully acknowledged for yearly FOE funding to the Euro-BioImaging Multi-Modal Molecular Imaging Italian Node (MMMI).

## Conflicts of interest

The authors declare that they have no conflict of interest.

## Availability of data and material

The datasets analyzed in the current study are openly available in the CIM-XNAT repository, http://cim-xnat.unito.it/. Users can access CIM-XNAT with the following credentials: username: xnat-pic-guest, password: preclinical. Upon login, you will be redirect to the XNAT webpage containing two projects related to the data presented in this work.

## Code availability (software application or custom code)

The latest releases of the source codes of XNAT-PIC are available to download from the GitHub repositories https://github.com/szullino/XNAT-PIC-Pipelines and https://github.com/szullino/XNAT-PIC. XNAT-PIC is a free software and is distributed under the terms of the GNU General Public License v3 or any later version as stated by the Free Software Foundation.

## Ethics approval

Not applicable

## Consent to participate

Not Applicable

## Consent for publication

Not Applicable

**Abbreviations**

| | |
|---|---|
| ADNI | Alzheimer's Disease Neuroimaging Initiative |
| AIM | Annotation Image Markup |
| API | Application Programming Interface |
| CEST | Chemical Exchange Saturation Transfer |
| CIM | Molecular Imaging Center |
| COINS | Collaborative Informatics and Neuroimaging Suite |
| CT | Computed Tomography |
| DICOM | Digital Imaging and COmmunications in Medicine |
| DICOM SEG | DICOM Segmentation |
| DOI | Digital Object Identifier |
| DS | Decimal String |
| DWI | Diffusion Weighted Imaging |
| EOSC | European Open Science Cloud |
| EPI2 | European Population Imaging Infrastructure |
| FAIR | Findable, Accessible, Interoperable and Reusable |
| FDG | 18-fluorodeoxyglucose |
| FLI | France Life Imaging |
| HID | Human Imaging Database |
| HTTP | Hypertext Transfer Protocol |
| LO | Long String |
| LORIS | Longitudinal Online Research and Imaging System |
| MIRMAID | Medical Imaging Research Management and Associated Information Database |
| MMMI | Multi-Modal Molecular Imaging Italian Node |
| MRI | Magnetic Resonance Imaging |
| NIfTI | Neuroimaging Informatics Technology Initiative |
| OASIS | Open Access Series of Imaging Studies |
| OCT | Optical Coherence Tomography |
| OHIF | Open Health Imaging Foundation |
| OI | Optical Imaging |
| OT | Other |
| PACS | Picture Archiving and Communication System |
| PAI | Photoacoustic Imaging |
| PET | Positron Emission Tomography |
| REST | Representational State Transfer |
| ROI | Region of Interest |
| RTSTRUCT | Radiotherapy Structure Set |
| SABER | Small Animal Big-data warehouse Environment for Research |
| SAS | Small Animal Shanoir |
| SCP | Service Class Provider |
| SPECT | Single Photon Emission Computed Tomography |
| TCIA | The Cancer Imaging Archive |
| VM | Value Multiplicity |
| VM | Value Multiplicity |
| VR | Value Representation |
| VR | Value Representation |
| XML | eXtensible Markup Language |
| XNAT | The Extensible Neuroimaging Archive Toolkit |
| XNAT-PIC | XNAT for Preclinical Imaging Center |

**References**

[1] F. Kiessling and B. J. Pichler, Eds., *Small Animal Imaging: Basics and Practical Guide*, 2nd Ed. Springer, 2017.

[2] J. K. Willmann, N. van Bruggen, L. M. Dinkelborg, and S. S. Gambhir, "Molecular imaging in drug development," *Nature Reviews Drug Discovery*, vol. 7, no. 7. Nature Publishing Group, pp. 591–607, 01-Jul-2008.

[3] J. Ellenberg *et al.*, "A call for public archives for biological image data," *Nature Methods*, vol. 15, no. 11. Nature Publishing Group, pp. 849–854, 01-Nov-2018.

[4] E. Williams *et al.*, "Image Data Resource: A bioimage data integration and publication platform," *Nat. Methods*, vol. 14, no. 8, pp. 775–781, Jul. 2017.

[5] S. Adebayo *et al.*, "PhenoImageShare: An image annotation and query infrastructure," *J. Biomed. Semantics*, vol. 7, no. 1, p. 35, Jun. 2016.

[6] "The Human Connectome Project." [Online]. Available: http://www.humanconnectomeproject.org/.

[7] K. Clark *et al.*, "The cancer imaging archive (TCIA): Maintaining and operating a public information repository," *J. Digit. Imaging*, vol. 26, no. 6, pp. 1045–1057, Dec. 2013.

[8] S. G. Mueller *et al.*, "The Alzheimer's disease neuroimaging initiative," *Neuroimaging Clinics of North America*, vol. 15, no. 4. Elsevier, pp. 869–877, 01-Nov-2005.

[9] D. S. Marcus, T. H. Wang, J. Parker, J. G. Csernansky, J. C. Morris, and R. L. Buckner, "Open Access Series of Imaging Studies (OASIS): Cross-sectional MRI data in young, middle aged, nondemented, and demented older adults," *J. Cogn. Neurosci.*, vol. 19, no. 9, pp. 1498–1507, Sep. 2007.

[10] S. Das, A. P. Zijdenbos, J. Harlap, D. Vins, and A. C. Evans, "LORIS: a web-based data management system for multi-center studies," *Front. Neuroinform.*, vol. 5, no. JANUARY 2012, p. 37, Jan. 2012.

[11] I. B. Ozyurt *et al.*, "Federated web-accessible clinical data management within an extensible neuroimaging database," *Neuroinformatics*, vol. 8, no. 4, pp. 231–249, Dec. 2010.

[12] A. Scott *et al.*, "Coins: An innovative informatics and neuroimaging tool suite built for large heterogeneous datasets," *Front. Neuroinform.*, vol. 5, Dec. 2011.

[13] P. D. Korfiatis, T. L. Kline, D. J. Blezek, S. G. Langer, W. J. Ryan, and B. J. Erickson, "MIRMAID: A Content Management System for Medical Image Analysis Research," *RadioGraphics*, vol. 35, no. 5, pp. 1461–1468, Sep. 2015.

[14] C. Anastasopoulos, M. Reisert, and E. Kellner, "'Nora Imaging': A Web-Based Platform for Medical Imaging," *Neuropediatrics*, vol. 48, no. S 01, pp. S1–S45, Apr. 2017.

[15] D. S. Marcus, T. R. Olsen, M. Ramaratnam, and R. L. Buckner, "The extensible neuroimaging archive toolkit: An informatics platform for managing, exploring, and sharing

neuroimaging data," *Neuroinformatics*, vol. 5, no. 1, pp. 11–33, 2007.

[16] "The Extensible Neuroimaging Archive Toolkit (XNAT)." [Online]. Available: http://xnat.org/.

[17] S. J. Doran *et al.*, "Informatics in Radiology: Development of a Research PACS for Analysis of Functional Imaging Data in Clinical Research and Clinical Trials," *RadioGraphics*, vol. 32, no. 7, pp. 2135–2150, 2012.

[18] T. Doel *et al.*, "GIFT-Cloud: A data sharing and collaboration platform for medical imaging research," 2016.

[19] J. Wu, C. Jansen, M. Beier, M. Witt, and D. Krefting, "Extending XNAT towards a Cloud-Based Quality Assessment Platform for Retinal Optical Coherence Tomographies," in *2014 14th IEEE/ACM International Symposium on Cluster, Cloud and Grid Computing*, 2014, pp. 764–773.

[20] M. Beier *et al.*, "Multicenter data sharing for collaboration in sleep medicine," *Futur. Gener. Comput. Syst.*, vol. 67, pp. 466–480, Feb. 2017.

[21] P. Kalendralis *et al.*, "Multicenter CT phantoms public dataset for radiomics reproducibility tests," *Med. Phys.*, vol. 46, no. 3, pp. 1512–1518, Mar. 2019.

[22] "The European Population Imaging Infrastructure (EPI2)." [Online]. Available: http://populationimaging.eu/.

[23] H. C. Achterberg, M. Koek, and W. J. Niessen, "Fastr: A Workflow Engine for Advanced Data Flows in Medical Image Analysis," *Front. ICT*, vol. 3, p. 15, Aug. 2016.

[24] "Health-RI XNAT." [Online]. Available: https://www.health-ri.nl/services/xnat.

[25] S. Klein, E. Vast, J. van Soest, A. Dekker, M. Koek, and W. Niessen, "XNAT imaging platform for BioMedBridges and CTMM TraIT," *J. Clin. Bioinforma.*, vol. 5, no. Suppl 1, p. S18, May 2015.

[26] M. Kain *et al.*, "Small Animal Shanoir (SAS) A Cloud-Based Solution for Managing Preclinical MR Brain Imaging Studies," *Front. Neuroinform.*, vol. 14, p. 20, May 2020.

[27] C. Barillot *et al.*, "Shanoir: Applying the Software as a Service Distribution Model to Manage Brain Imaging Research Repositories," *Front. ICT*, vol. 3, no. DEC, p. 25, Dec. 2016.

[28] "Digital Imaging and COmmunications in Medicine (DICOM)." [Online]. Available: https://www.dicomstandard.org/.

[29] "Neuroimaging Informatics Technology Initiative (NIfTI)." [Online]. Available: https://nifti.nimh.nih.gov/.

[30] P. Mildenberger, M. Eichelberg, and E. Martin, "Introduction to the DICOM standard," *European Radiology*, vol. 12, no. 4. Springer, pp. 920–927, 01-Apr-2002.

[31] "Python Software Foundation." [Online]. Available: https://www.python.org/.

[32] S. van der Walt, S. C. Colbert, and G. Varoquaux, "The NumPy Array: A Structure for Efficient Numerical Computation," *Comput. Sci. Eng.*, vol. 13, no. 2, pp. 22–30, Mar. 2011.

[33] "numpy · PyPI." [Online]. Available: https://pypi.org/project/numpy/1.15.4/. [Accessed: 18-Feb-2021].

[34] D. Mason, "SU-E-T-33: Pydicom: An Open Source DICOM Library," in *Medical Physics*, 2011, vol. 38, no. 6, p. 3493.

[35] "pydicom · PyPI." [Online]. Available: https://pypi.org/project/pydicom/1.2.1/. [Accessed: 18-

neuroimaging data," *Neuroinformatics*, vol. 5, no. 1, pp. 11–33, 2007.

[16] "The Extensible Neuroimaging Archive Toolkit (XNAT)." [Online]. Available: http://xnat.org/.

[17] S. J. Doran *et al.*, "Informatics in Radiology: Development of a Research PACS for Analysis of Functional Imaging Data in Clinical Research and Clinical Trials," *RadioGraphics*, vol. 32, no. 7, pp. 2135–2150, 2012.

[18] T. Doel *et al.*, "GIFT-Cloud: A data sharing and collaboration platform for medical imaging research," 2016.

[19] J. Wu, C. Jansen, M. Beier, M. Witt, and D. Krefting, "Extending XNAT towards a Cloud-Based Quality Assessment Platform for Retinal Optical Coherence Tomographies," in *2014 14th IEEE/ACM International Symposium on Cluster, Cloud and Grid Computing*, 2014, pp. 764–773.

[20] M. Beier *et al.*, "Multicenter data sharing for collaboration in sleep medicine," *Futur. Gener. Comput. Syst.*, vol. 67, pp. 466–480, Feb. 2017.

[21] P. Kalendralis *et al.*, "Multicenter CT phantoms public dataset for radiomics reproducibility tests," *Med. Phys.*, vol. 46, no. 3, pp. 1512–1518, Mar. 2019.

[22] "The European Population Imaging Infrastructure (EPI2)." [Online]. Available: http://populationimaging.eu/.

[23] H. C. Achterberg, M. Koek, and W. J. Niessen, "Fastr: A Workflow Engine for Advanced Data Flows in Medical Image Analysis," *Front. ICT*, vol. 3, p. 15, Aug. 2016.

[24] "Health-RI XNAT." [Online]. Available: https://www.health-ri.nl/services/xnat.

[25] S. Klein, E. Vast, J. van Soest, A. Dekker, M. Koek, and W. Niessen, "XNAT imaging platform for BioMedBridges and CTMM TraIT," *J. Clin. Bioinforma.*, vol. 5, no. Suppl 1, p. S18, May 2015.

[26] M. Kain *et al.*, "Small Animal Shanoir (SAS) A Cloud-Based Solution for Managing Preclinical MR Brain Imaging Studies," *Front. Neuroinform.*, vol. 14, p. 20, May 2020.

[27] C. Barillot *et al.*, "Shanoir: Applying the Software as a Service Distribution Model to Manage Brain Imaging Research Repositories," *Front. ICT*, vol. 3, no. DEC, p. 25, Dec. 2016.

[28] "Digital Imaging and COmmunications in Medicine (DICOM)." [Online]. Available: https://www.dicomstandard.org/.

[29] "Neuroimaging Informatics Technology Initiative (NIfTI)." [Online]. Available: https://nifti.nimh.nih.gov/.

[30] P. Mildenberger, M. Eichelberg, and E. Martin, "Introduction to the DICOM standard," *European Radiology*, vol. 12, no. 4. Springer, pp. 920–927, 01-Apr-2002.

[31] "Python Software Foundation." [Online]. Available: https://www.python.org/.

[32] S. van der Walt, S. C. Colbert, and G. Varoquaux, "The NumPy Array: A Structure for Efficient Numerical Computation," *Comput. Sci. Eng.*, vol. 13, no. 2, pp. 22–30, Mar. 2011.

[33] "numpy · PyPI." [Online]. Available: https://pypi.org/project/numpy/1.15.4/. [Accessed: 18-Feb-2021].

[34] D. Mason, "SU-E-T-33: Pydicom: An Open Source DICOM Library," in *Medical Physics*, 2011, vol. 38, no. 6, p. 3493.

[35] "pydicom · PyPI." [Online]. Available: https://pypi.org/project/pydicom/1.2.1/. [Accessed: 18-

Feb-2021].

[36] M. Caffini, "Project-Beat--Pyhton." [Online]. Available: https://github.com/mcaffini/Project-Beat---Python.

[37] "xnat · PyPI." [Online]. Available: https://pypi.org/project/xnat/0.3.22/. [Accessed: 18-Feb-2021].

[38] "pyAesCrypt · PyPI." [Online]. Available: https://pypi.org/project/pyAesCrypt/0.4.3/. [Accessed: 18-Feb-2021].

[39] "pyinstaller · PyPI." [Online]. Available: https://pypi.org/project/pyinstaller/3.5/. [Accessed: 18-Feb-2021].

[40] "Get Started with MATLAB Engine API for Python - MATLAB & Simulink." [Online]. Available: https://www.mathworks.com/help/matlab/matlab_external/get-started-with-matlab-engine-for-python.html. [Accessed: 18-Feb-2021].

[41] Y. Schwartz et al., "PyXNAT: XNAT in Python," *Front. Neuroinform.*, vol. 6, p. 12, May 2012.

[42] "pyxnat · PyPI." [Online]. Available: https://pypi.org/project/pyxnat/1.2.1.0.post3/. [Accessed: 18-Feb-2021].

[43] "requests · PyPI." [Online]. Available: https://pypi.org/project/requests/2.23.0/. [Accessed: 18-Feb-2021].

[44] "numpy · PyPI." [Online]. Available: https://pypi.org/project/numpy/1.18.5/. [Accessed: 18-Feb-2021].

[45] T. Phil, "Sikerdebaard/dcmrtstruct2nii: v1.0.19," 19-Sep-2020. [Online]. Available: https://zenodo.org/record/4037865. [Accessed: 18-Feb-2021].

[46] "opencv-python · PyPI." [Online]. Available: https://pypi.org/project/opencv-python/4.4.0.40/. [Accessed: 18-Feb-2021].

[47] M. Brett et al., "nipy/nibabel: 3.1.1," 30-Jun-2020. [Online]. Available: https://zenodo.org/record/3924343. [Accessed: 18-Feb-2021].

[48] L. Consolino et al., "Non-invasive Investigation of Tumor Metabolism and Acidosis by MRI-CEST Imaging," *Frontiers in Oncology*, vol. 10. Frontiers Media S.A., p. 161, 18-Feb-2020.

[49] O. S. Pianykh, *Digital imaging and communications in medicine (DICOM): A practical introduction and survival guide*. Berlin, Heidelberg: Springer Berlin Heidelberg, 2008.

[50] D. L. Longo et al., "In vivo imaging of tumor metabolism and acidosis by combining PET and MRI-CEST pH imaging," *Cancer Res.*, vol. 76, no. 22, pp. 6463–6470, 2016.

[51] K. W. Y. Chan et al., "Natural D -glucose as a biodegradable MRI contrast agent for detecting cancer," *Magn. Reson. Med.*, vol. 68, no. 6, pp. 1764–1773, Dec. 2012.

[52] S. Walker-Samuel et al., "In vivo imaging of glucose uptake and metabolism in tumors," *Nat. Med.*, vol. 19, no. 8, pp. 1067–1072, 2013.

[53] "XNAT OHIF Viewer 2.0 Plugin." [Online]. Available: https://bitbucket.org/icrimaginginformatics/ohif-viewer-xnat-plugin/src/master/.

[54] T. Urban et al., "LesionTracker: Extensible open-source zero-footprint web viewer for cancer imaging research and clinical trials," *Cancer Res.*, vol. 77, no. 21, pp. e119–e122, Nov. 2017.

[55] "dcmrtstruct2ni: Convert DICOM RT-Struct to nii." [Online]. Available:

https://pypi.org/project/dcmrtstruct2nii/.

[56] "DICOM Modality." [Online]. Available: https://www.dicomlibrary.com/dicom/modality/.

[57] L. Persoon *et al.*, "A novel data management platform to improve image-guided precision preclinical biological research," *Br. J. Radiol.*, vol. 92, no. 1095, 2019.

[58] "Flywheel • Informatics Platform for Biomedical Research & Collaboration." [Online]. Available: https://flywheel.io/. [Accessed: 17-Feb-2021].

[59] M. D. Wilkinson *et al.*, "The FAIR Guiding Principles for scientific data management and stewardship," *Sci. Data*, vol. 3, p. 160018, Mar. 2016.

[60] P. Jansen, L. van den Berg, P. van Overveld, and J. W. Boiten, "Research data stewardship for healthcare professionals," in *Fundamentals of Clinical Data Science*, Springer International Publishing, 2018, pp. 37–53.

[61] "Euro-BioImaging ERIC." [Online]. Available: https://www.eurobioimaging.eu/.

[62] "The Vienna Declaration on the European Open Science Cloud." [Online]. Available: https://eosc-launch.eu/declaration/.